\documentclass[preprint,showpacs,preprintnumbers,amsmath,amssymb]{revtex4}

\usepackage{graphicx}% Include figure files
\usepackage{dcolumn}% Align table columns on decimal point
\usepackage{bm}% bold math

\begin{document}

%\preprint{APS/123-QED}

\title{Topology of \lq\lq white\rq\rq stars in relativistic fragmentation of light nuclei}

\author{N.~P.~Andreeva}
   \affiliation{Institute of Physics and Technology, Almaty, Kazakstan}
\author{V.~Bradnova}
   \affiliation{Joint Insitute for Nuclear Research, Dubna, Russia} 
\author{S.~Vok\'al}
   \affiliation{P. J. \u Saf\u arik University, Ko\u sice, Slovak Republic}
\author{A.~Vok\'alov\'a}
   \affiliation{Joint Insitute for Nuclear Research, Dubna, Russia} 
\author{A.~Sh.~Gaitinov}
   \affiliation{Institute of Physics and Technology, Almaty, Kazakstan} 
\author{S.~G.~Gerasimov}
   \affiliation{Lebedev Institute of Physics, Russian Academy of Sciences, Moscow, Russia} 
\author{L.~A.~Goncharova}
   \affiliation{Lebedev Institute of Physics, Russian Academy of Sciences, Moscow, Russia} 
\author{V.~A.~Dronov}
   \affiliation{Lebedev Institute of Physics, Russian Academy of Sciences, Moscow, Russia}
\author{P.~I.~Zarubin}
     \email{zarubin@lhe.jinr.ru}    
     \homepage{http://becquerel.lhe.jinr.ru}
     \homepage{http://pavel.jinr.ru}
   \affiliation{Joint Insitute for Nuclear Research, Dubna, Russia} 
 \author{I.~G.~Zarubina}
   \affiliation{Joint Insitute for Nuclear Research, Dubna, Russia}    
\author{G.~I.~Orlova}
   \affiliation{Lebedev Institute of Physics, Russian Academy of Sciences, Moscow, Russia} 
\author{A.~D.~Kovalenko}
   \affiliation{Joint Insitute for Nuclear Research, Dubna, Russia}  
\author{A.~Kravchakova}
   \affiliation{P. J. \u Saf\u arik University, Ko\u sice, Slovak Republic}
\author{V.~G.~Larionova}
   \affiliation{Lebedev Institute of Physics, Russian Academy of Sciences, Moscow, Russia} 
\author{F.~G.~Lepekhin}
   \affiliation{Institute of Nuclear Physics, Russian Academy of Sciences, Gatchina, Russia} 
\author{O.~V.~Levitskaya}
   \affiliation{Institute of Nuclear Physics, Russian Academy of Sciences, Gatchina, Russia} 
\author{A.~I.~Malakhov}
   \affiliation{Joint Insitute for Nuclear Research, Dubna, Russia} 
\author{A.~A.~Moiseenko}
   \affiliation{Institute of Physics, Erevan, Armenia}
\author{G.~I.~Orlova}
   \affiliation{Lebedev Institute of Physics, Russian Academy of Sciences, Moscow, Russia} 
\author{N.~G.~Peresadko}
   \affiliation{Lebedev Institute of Physics, Russian Academy of Sciences, Moscow, Russia} 
\author{N.~G.~Polukhina}
   \affiliation{Lebedev Institute of Physics, Russian Academy of Sciences, Moscow, Russia} 
\author{P.~A.~Rukoyatkin}
   \affiliation{Joint Insitute for Nuclear Research, Dubna, Russia} 
\author{V.~V.~Rusakova}
   \affiliation{Joint Insitute for Nuclear Research, Dubna, Russia} 
\author{N.~A.~Salmanova}
  \affiliation{Lebedev Institute of Physics, Russian Academy of Sciences, Moscow, Russia} 
\author{V.~R.~Sarkisyan}
   \affiliation{Institute of Physics, Erevan, Armenia}
\author{B.~B.~Simonov}
   \affiliation{Institute of Nuclear Physics, Russian Academy of Sciences, Gatchina, Russia}
\author{E.~Stan}
   \affiliation{Institute of Space Sciences, Magurele, Romania}   
\author{R.~Stanoeva}
   \affiliation{Institute of Nuclear Researches and Nuclear Energy, Bulgarian Academy of Sciences, Bulgaria}       
\author{M.~M.~Chernyavsky}
  \affiliation{Lebedev Institute of Physics, Russian Academy of Sciences, Moscow, Russia} 
\author{M.~Haiduc}
   \affiliation{Institute of Space Sciences, Magurele, Romania}     
\author{S.~P.~Kharlamov}
   \affiliation{Lebedev Institute of Physics, Russian Academy of Sciences, Moscow, Russia}
\author{I.~Tsakov}
   \affiliation{Institute of Nuclear Researches and Nuclear Energy, Bulgarian Academy of Sciences, Bulgaria}    
\author{T.~V.~Shchedrina}
   \affiliation{Joint Insitute for Nuclear Research, Dubna, Russia}

\date{\today}% It is always \today, today,

\begin{abstract}
\indent   In the present paper, experimental observations of the multifragmentation processes of light 
relativistic nuclei carried out by means of emulsions are reviewed. Events of the type of 
\lq\lq white\rq\rq ~stars in which the dissociation of relativistic nuclei is not accompanied by the production of mesons and the
 target-nucleus fragments are considered.\par
\indent  A distinctive feature of the charge topology in the dissociation of the Ne, Mg, Si, and S nuclei is an almost 
total suppression of the binary splitting of nuclei to fragments with charges higher than 2. The growth of the nuclear
 fragmentation degree is revealed in an increase in the multiplicity of singly and doubly charged fragments with 
decreasing charge of the non-excited part of the fragmenting nucleus.          
\par
\indent The processes of dissociation of stable Li, Be, B, C, N, and O isotopes to charged fragments were used to
 study special features of the formation of systems consisting of the lightest  $\alpha$, d, and t nuclei. Clustering in 
form of the $^3$He nucleus can be detected in \lq\lq white\rq\rq ~stars via the dissociation of neutron-deficient Be, 
B, C, and N isotopes.  
\par
\end{abstract}
  %    {PACS-key}{21.45.+v} \and
   %   {PACS-key}{23.60+e} \and
    %  {PACS-key}{25.10.+s}  

\maketitle
\section{\label{sec:level1}Introduction}

\indent  The charge topology of fragments in peripheral interactions of light nuclei of an initial energy above
1 GeV per nucleon may be considered as an effective characteristic of the nuclear multifragmentation phenomenon. 
 In this energy range, a regime of limiting fragmentation of nuclei sets in, that is, the fragment spectrum is 
invariable with respect to the collision energy and the target-nucleus composition. 
\par
\indent  In the investigation of the multifragmentation at relativistic energies, the possibilities of observing
 the final states consisting of charged fragments and their spectroscopy are defined by the accuracy of angular 
measurements. Owing to the best spatial resolution (0.5 $\mu$m), the nuclear emulsion ensures the angular resolution 
of the tracks of relativistic fragments of about 10$^{-5}$ rad. This enables one to observe completely all the possible 
decays of nuclear excited states to fragments. For example, over a track length of 1 mm, one can surely distinguish a 
process $^8$Be$\rightarrow$2$\alpha$, which is revealed for a momentum of 4.5 GeV/c per nucleon as a pair of tracks 
within an angular cone of about 2$\cdot$10$^{-3}$  rad. Such narrow decays are rather frequently observed in the
 fragmentation of relativistic oxygen nuclei, as well as heavier ones.    
\par
\indent The topologic characteristics of the events in the dissociation of light nuclei in peripheral interactions
 were investigated by the emulsion technique for the following nuclei   $^{12}$C \cite{Adamovich77,Marin80,
Abdurazakova84,Belaga95,Belaga95a,Bondarenko98}, $^{22}$Ne \cite{Andreeva85,Karabova86,
Andreeva88,Andreeva88a,Naghy88,Naghy85}, 
$^{24}$Mg \cite{Bondarenko91},$^{28}$Si \cite{Krasnov88,Ameeva90,Adamovich92}, 
$^{16}$O \cite{Adamovich92a,Avetyan96}, $^{6}$Li \cite{Lepekhin95,Lepekhin98,Adamovich99,Adamovich03}, and $^{10}$B 
\cite{Bradnova03,Bradnova03a,Adamovich04}  at energies of the order of a few GeV per nucleon. The dissociation of 
the $^{16}$O and $^{32}$S nuclei at an energy of 200 GeV per nucleon was studied in 
\cite{Adamovich92a,Baroni90,Baroni92}.  All these results are notable
for their exceptional completeness and reliability. They may turn out to be useful for planning investigations on nuclear
 multifragmentation with a high statistical provision.  
\par
\indent The present paper gives the data on the dissociation channels for a wide range  of light nuclei  in events 
of the \lq\lq white\rq\rq ~star  type. The experimental data on the relations between the nuclear dissociation channels 
being observed give an idea both of the general features of nuclear fragmentation processes and the special ones associated
 with the structure of individual nuclei. The results for the $^{24}$Mg, $^{14}$N, and $^{7}$Be nuclei are presented for 
the first time. The results concerning other nuclei in question were obtained by using the events from earlier published
data that were selected on the basis of more rigid criteria. Emulsions were exposed to beams of energy of a few Gev per 
nucleon at the JINR Synchrophasotron and Nuclotron, while to beams of energy of 200 GeV per nucleon - in CERN.  
\par

\begin{figure}
\includegraphics[width=160mm]{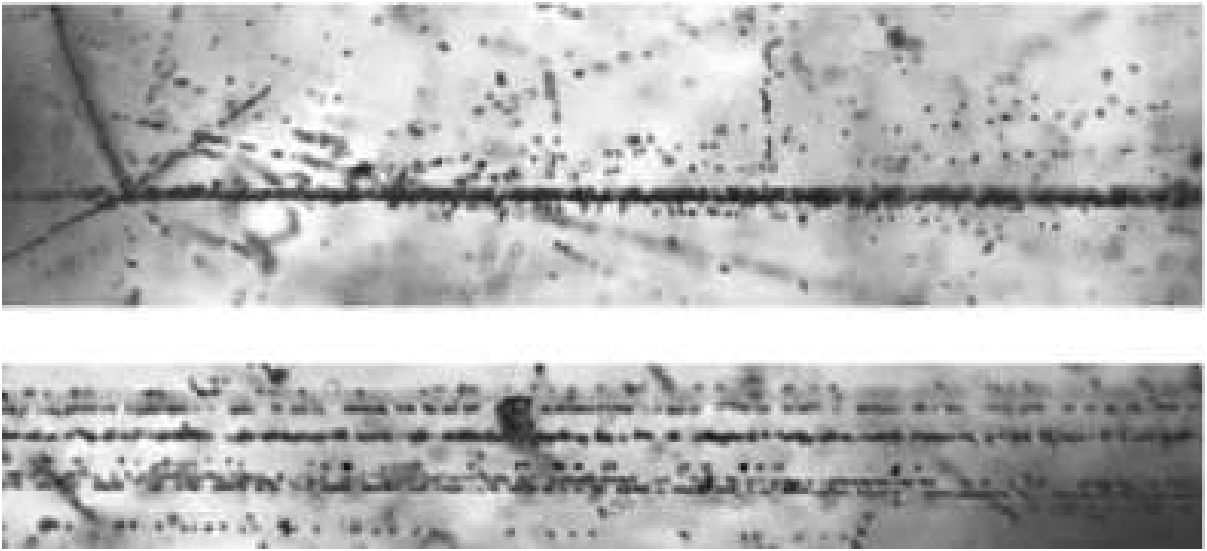} 
\caption{\label{fig:1}Event of the fragmentation of a $^{28}$Si nucleus of the energy of 3.65 GeV per nucleon in a 
peripheral interaction on an emulsion nucleus. On the upper photograph one can see the interaction vertex
  and the jet of fragments in a narrow angular cone along with four accompanying singly charged particles
 in a wide cone and three fragments of the target-nucleus. Moving toward the fragment jet direction  
(upper photograph) it is possible to distinguish 3 fragments of hydrogen and 5 fragments of helium. An
 intensive track on the upper photograph  (the third one from above) is identified as a very narrow pair
  of Z=2 fragments  corresponding   to the $^{8}$Be decay. A 3-dimensional image of the event was reconstructed as 
a plane projection by means of an automated microscope (Lebedev Institute of Physics, Moscow) of the PAVIKOM complex.}
\end{figure}

\indent For the sake of illustration of the selection criteria, Figure \ref{fig:1} presents an event of the multifragmentation of a 
$^{28}$Si nucleus of momentum 4.5 GeV/c per nucleon. Of particular   interest is a group of projectile fragments with
 total charge Z=13 within a narrow cone of angles of the order of a few degrees. The magnitude of the cone is defined 
by the ratio of the transverse Fermi momentum to the momentum per nucleon of the primary nucleus. The tracks of relativistic
fragments remain in one emulsion layer sufficiently for reconstructing a continuous three-dimensional image of this 
group of tracks. The mass identification of relativistic H and He isotopes in emulsion is possible via the determination 
of the mean angle of track scattering and the total momentum connected with it.  
\par
\indent The longitudinal momenta of fragments per nucleon are equal, within a few percent, to the momenta of the nucleons 
of the primary nucleus. The excitation energy of a system of fragments is defined by their multiplicity and the emission 
angles. It can be estimated as the difference between the invariant mass of the fragmenting system and the mass of the 
primary nucleus and amounts to a few MeV per nucleon of the fragment. The angular correlations of fragments reflect the 
angular momentum of the produced system. In Figure \ref{fig:2}, one can see, in a broader cone, tracks with minimal ionization from 
produced mesons. In addition, in the interaction peak, there are tracks from strongly ionizing target-nucleus fragments
 of energy of the order of several tens of MeV. Thus, the separation of the kinematic regions of the fragmentation of 
colliding nuclei is clearly revealed in the interaction. 
\par
 \subsection{Multifragmentaion in \lq\lq white\rq\rq ~ stars}
\indent When accumulating data on nuclear multifragmentation,
 events without tracks from charged particles are selected between the areas of the fragmentation of a projectile and 
the target-nucleus. As a rule, in such events the primary nucleus charge is totally transferred into a narrow angular cone 
of fragmentation. The most obvious interpretation is provided for the events, which contain no tracks also from the
 target-nucleus fragments. They are produced in the case of a minimal energy transfer to the fragmenting nucleus. Events 
of such a type are called \lq\lq white\rq\rq ~stars because of their appearance. Their fraction constitutes few percent 
of the total number of inelastic events. Their name reflects not only the outward look of the event, but also a sharp
 decrease of ionization losses (in a limiting case, by a factor of Z) in the transition from the primary nucleus track 
to the narrow cone of secondary tracks. The formation of \lq\lq white\rq\rq ~stars is induced by the electromagnetic 
interactions of  the target-nucleus with virtual photons and by the diffraction scattering on peripheral  target neutrons. 
\par
\indent In the search for events of this type, of important practical advantage is the requirement of charge conservation,
 which makes it possible to exclude in the beam admixtures from lighter nuclei with a close charge-to-mass ratio. 
This condition is essential when emulsion is exposed to the secondary beams of radioactive nuclei having a rather
 complicated composition. We note that the above-mentioned criteria of selection of \lq\lq white\rq\rq ~stars along with the 
requirement of conservation of the energy flux in the fragmentation cone can be used in a future experiment dealing 
with the study of global features of the fragmentation of heavy nuclei in peripheral dissociation processes.   
\par
   \subsection{Loosely bound cluster systems} 
\indent The goal of our experiments is the study of the picture of the phase 
transition of nuclear matter from the state of a quantum liquid to that of a quantum gas consisting of a large number
 of nucleons and the lightest nuclei that occurs near the energy thresholds. The term \lq\lq lightest nuclei\rq\rq ~implies
 deuterons and tritons, as well as $^3$He and $^3$He nuclei, that is, stable systems having no excited states below nucleon 
decay thresholds.   
\par
\indent    The present-day interest in the study of such phase transitions is motivated by the prediction of the
 properties of such states as loosely bound cluster systems \cite{Efimov70,Nunes03}. The spatial extension of these systems can essentially
 exceed the sizes of the fragments (Efimov states \cite{Efimov70} near the threshold of the decay of 3-body systems, light nuclei
 having the structure of a molecular type [29], the Bose  condensate of dilute $\alpha$ particle gas in N$\alpha$
 nuclei \cite{Schuck03}). A multifragmentation process going with an adiabatic transfer of excitation and without nucleon exchange 
may be interpreted as a disappearance of the Coulomb barrier because of a simultaneous increase in distances between 
charged clusters.  
\par
\indent The study of such states on the scale typical of the nucleon and cluster structure of the nucleus is of interest
 for nuclear astrophysics. For example, thanks to an essential decrease in the Coulomb repulsion in such extended systems,
the latter can play the role of intermediate states in nucleosynthesis processes in stars. The topologies established can
 turn out to be useful for clearing up the variants of the nuclear synthesis as processes inverse to those of their
 fragmentation.
\par
\section{\label{sec:level2} PARTICULAR FEATURES OF THE FRAGMENTATION OF
Ne, Mg, Si, and S NUCLEI}
 \subsection{Multifragmentation of $^{24}$Mg nuclei}

 \indent  \lq\lq White\rq\rq ~stars were sought in the dissociation of the
 $^{24}$Mg nuclei of a kinetic energy of 3.65 GeV per nucleon by viewing by means of microscopes along the primary n
ucleus track up to the interaction peaks (e.g., \cite{Adamovich04}). 83 events of this type were found in which almost all 
secondary tracks were confined within a 4$^\circ$ angular cone  to the primary track direction. The value of the charge of
a particle forming the track in emulsion was estimated by the density of ruptures on the track and the number of $\delta$
 electrons. The distribution of the events with respect to the charge topology of fragments is given in table 1. 
The upper row is the charge of a fragment with Z$>$2, the second row the number of singly-charged fragments, 
the third one - the number of doubly-charged fragments, and the bottom one the number of the events found with 
this topology. The observation of events with an 11+1 topology was effective with a level of about 50 \% since the singly
 charged tracks were screened by the second track of a large ionization.  
\par

\begin{table}
\caption{\label{tab:1}  The distribution of 
\lq\lq white\rq\rq ~stars with respect to the charge topology in dissociation of $^{24}$Mg of 
the energy of 3.65 GeV per nucleon. }
\begin{tabular}{c|c|c|c|c|c|c|c|c|c|c}

\hline\noalign{\smallskip}
\hline\noalign{\smallskip}

~Z$_f$~	   &~~ 11 ~&~~ 10 ~&~~ 10 ~&~~ 9 ~&~~ 9 ~&~~ 8 ~&~~ 8 ~&~~  8  ~&~~ 7 ~&~~ 7 ~\\
~N$_{Z=1}$~&~~  1 ~&~~  2 ~&~~    ~&~~ 3 ~&~~ 1 ~&~~ 4 ~&~~ 2 ~&~~    ~~&~~ 3 ~&~~ 1 ~\\
~N$_{Z=2}$~&~~    ~&~~    ~&~~  1 ~&~~   ~&~~ 1 ~&~~   ~&~~ 1 ~&~~  2 ~~&~~ 1 ~&~~ 2 ~\\
~N$_{ev}$~ &~~ 10 ~&~  14 ~&~~  8 ~&~~ 5 ~&~~ 9 ~&~~ 1 ~&~~ 7 ~&~~  4 ~~&~~ 4 ~&~~ 2 ~\\

\hline\noalign{\smallskip}
\hline\noalign{\smallskip}
\end{tabular}
\begin{tabular}{c|c|c|c|c|c|c|c|c|c|c}

\hline\noalign{\smallskip}
\hline\noalign{\smallskip}

~Z$_f$~	   &~~  6 ~&~~  5 ~&~~  5 ~&~~ 5 ~&~~ 4 ~&~~ 4 ~&~~ 3 ~&~~     ~&~~   ~&~~   ~\\
~N$_{Z=1}$~&~~  2 ~&~~  5 ~&~~  3 ~&~~ 1 ~&~~ 6 ~&~~ 4 ~&~~ 5 ~&~~  6 ~~&~~ 4 ~&~~ 2 ~\\
~N$_{Z=2}$~&~~  2 ~&~~  1 ~&~~  2 ~&~~ 3 ~&~~ 1 ~&~~ 2 ~&~~ 2 ~&~~  3 ~~&~~ 4 ~&~~ 5 ~\\
~N$_{ev}$~ &~~  4 ~&~   2 ~&~~  1 ~&~~ 1 ~&~~ 2 ~&~~ 1 ~&~~ 3 ~&~~  1 ~~&~~ 2 ~&~~ 2 ~\\

\hline\noalign{\smallskip}
\hline\noalign{\smallskip}
\end{tabular}
\end{table}
\indent Table \ref{tab:1} contains the production channels for \lq\lq white\rq\rq ~stars starting with the separation of individual
singly and doubly charged fragments from a  \lq\lq cold remainder\rq\rq ~of the primary nucleus to its total   breakup 
into lightest nuclei. In no one of these events there are more than one track from the relativistic fragments with Z$>$2. 
The obvious particular feature is the absence of the events of binary and triple splitting of light nuclei to fragments 
heavier than the alpha particle, which suggests the dominance of the contribution from the multifragmentation process.
Earlier, Mg$^*\rightarrow$B+N splitting alone C without an additional emission of charged particles was observed in 
an analysis of 1666 interactions \cite{Bogdanov86}. Thus, the multifragmentation processes are dominating, in spite of their higher thresholds.
This fact is can be explained by a high density of multi-particle states.    
\par

\indent  It is planned to analyze the events due to a total Mg breakup using much more information with identification of
 lightest nuclei. In so doing, there will arise a possibility of reconstructing the invariant mass of a system being
 disintegrated and its subsystems (e.g., N$\alpha$ ones). By the present time, two events that are due to the decay of 
a $^{24}$Mg nucleus into six He nuclei are found. One of them is identified as 5$^{4}$He+$^{3}$He, they do not enter the
 statistics of Table \ref{tab:1} as far as these events are accompanied by single fragments of the target nucleus. Nevertheless, 
they give sufficient grounds to pursue further search for the 6$\alpha$ configurations over larger lengths of the primary 
tracks of $^{24}$Mg nuclei.
\par
\begin{table}
\caption{\label{tab:2}  The distribution of 
\lq\lq white\rq\rq ~stars with respect to the charge topology in dissociation of $^{22}$Ne of 
the energy of 3.27 GeV per nucleon. }
\begin{tabular}{c|c|c|c|c|c|c|c|c|c|c|c|c}

\hline\noalign{\smallskip}
\hline\noalign{\smallskip}

~Z$_f$~	   &~~ 9 ~&~~ 8 ~&~~ 8 ~&~~ 7 ~&~~ 6 ~&~~ 5 ~&~~ 5 ~&~ 5+3 ~&~~ 4 ~&~ 4+3 ~&~~   ~&~~   ~\\
~N$_{Z=1}$~&~~ 1 ~&~~   ~&~~ 2 ~&~~ 1 ~&~~ 2 ~&~~ 1 ~&~~ 3 ~&~~   ~~&~~   ~&~~  3 ~&~~ 2 ~&~~   ~\\
~N$_{Z=2}$~&~~   ~&~~ 1 ~&~~   ~&~~ 1 ~&~~ 1 ~&~~ 2 ~&~~ 1 ~&~~ 1 ~~&~~ 3 ~&~~    ~&~~ 4 ~&~~ 5 ~\\
~N$_{ev}$~ &~ 22 ~&~ 51 ~&~~ 6 ~&~~ 7 ~&~~ 2 ~&~~ 1 ~&~~ 1 ~&~~ 1 ~~&~~ 2 ~&~~  1 ~&~~ 1 ~&~~ 3 ~\\

\hline\noalign{\smallskip}
\hline\noalign{\smallskip}
\end{tabular}
\end{table}
\subsection{ Multifragmentation of $^{22}$Ne nuclei} 
\indent We compare the particular features of the $^{24}$Mg nucleus 
fragmentation with a large amount of information on the interactions of neighboring nuclei. Table \ref{tab:2} shows the charge 
topology distribution for 103 \lq\lq white\rq\rq ~stars originating from $^{22}$Ne nuclei with the energy 3.27 GeV per 
nucleon that were selected from 4100 inelastic events \cite{Andreeva85}. In this case as well, there are no binary splitting events. 
In \cite{Bogdanov86}, using some other statistic set from 4155 events, binary $^{22}$Ne splittings were   also not observed.
\par
\indent  A noticeably more evident role of the helium isotopes in the $^{22}$Ne fragmentation may be associated with 
the fact that, contrary to the symmetric Mg nucleus, the $^{22}$Ne nucleus has a pair of additional external neutrons. 
This situation can be employed for more effective generation and detection of systems consisting of a large number of $\alpha$
 particles in initiating dissociation via a knockout of external neutrons. We found three events from the decay of 
$^{22}$Ne nuclei to five He nuclei (Table \ref{tab:2}) the tracks of which were confined within a cone of 3$^\circ$. Of them, in 
two events, all the tracks were even within 1$^\circ$. These discoveries confirm once more unique capabilities of nuclear
 emulsions as applied to the investigation of multiparticle systems consisting of the lightest nuclei with minimal relative
4-velocities  (or relative Lorentz factors).   
\par
\subsection{Multifragmentation $^{28}$Si and $^{32}$S nuclei}
 \indent A statistic set of 116 \lq\lq white\rq\rq ~stars from the
 $^{28}$Si nuclei of the energy of 3.65 GeV per nucleon demonstrates the same particular feature, that is, the transition 
to the multifragmentation  (Table \ref{tab:3}) occurs by avoiding a binary splitting \cite{Krasnov88}. In \cite{Bogdanov86}, using another sample
from 1900 inelastic interactions, one observed an event Si$^*\rightarrow$O+C alone. It is interesting to note that the 
transition   to a total break up of a $^{28}$Si nucleus proceeds with an increasing contribution to the final states from 
the H isotopes with respect to the He isotopes. It has to be decided whether this fact is a consequence of a weakening of
the alpha clustering in nuclei with increasing A. The results mentioned represent an event sample from earlier obtained
 data \cite{Andreeva85,Krasnov88}. At the same time, these papers contain rich information, which is useful for planning experiments with 
varying inelasticity of selected collisions.  
\par

\begin{table}
\caption{\label{tab:3}  The distribution of 
\lq\lq white\rq\rq ~stars with respect to the charge topology in dissociation of $^{28}$Si of 
the energy of 3.65 GeV per nucleon. }

\begin{tabular}{c|c|c|c|c|c|c|c|c|c|c|c|c|c|c}

\hline\noalign{\smallskip}
\hline\noalign{\smallskip}

~~~Z$_f$~~~	     &  ~~13~~&	~~12~~&	~~12~~&  ~~11~~& ~~11~~&  ~~10~~& ~~10~~& ~~10~~& ~~~9~~& ~~~9~~& ~~~9~~ &  ~~~8~~&	~~~8~~&	~~~8~~\\
~~~N$_{Z=1}$~~~	 &  ~~~1~~& ~~~ ~~&	~~~2~~&  ~~~1~~& ~~~3~~&  ~~~ ~~& ~~~2~~& ~~~4~~& ~~~1~~& ~~~3~~& ~~~5~~&  ~~~6~~& ~~~2~~&	~~~4~~\\
~~~N$_{Z=2}$~~~	 &  ~~~ ~~&	~~~1~~&	~~~ ~~&  ~~~1~~& ~~~ ~~&  ~~~2~~& ~~~1~~& ~~~ ~~& ~~~2~~& ~~~1~~& ~~~ ~~&  ~~~ ~~&	~~~2~~&	~~~1~~\\
~~~N$_{ev}$~~~	 &  ~~~9~~&	~~ 3~~& ~~15~~&  ~~11~~& ~~~6~~&  ~~~2~~& ~~~7~~& ~~~2~~& ~~~2~~& ~~~8~~& ~~~3~~&  ~~~2~~&	~~~5~~& ~~~6~~\\

\hline\noalign{\smallskip}
\hline\noalign{\smallskip}
\end{tabular}

\begin{tabular}{c|c|c|c|c|c|c|c|c|c|c|c|c|c}

\hline\noalign{\smallskip}
\hline\noalign{\smallskip}

~~~Z$_f$~~~	     &  ~~~7~~& ~~~7~~&  ~~~7~~& ~~~6~~& ~~~6~~& ~~~6~~& ~~~6~~& ~~~5~~& ~~~5~~& ~~~4~~& ~~~ ~~& ~~~ ~~& ~~~ ~~\\
~~~N$_{Z=1}$~~~	 &  ~~~3~~& ~~~5~~&  ~~~7~~& ~~~2~~& ~~~4~~& ~~~6~~& ~~~8~~& ~~~3~~& ~~~5~~& ~~~2~~& ~~~2~~& ~~~8~~& ~~10~~\\
~~~N$_{Z=2}$~~~	 &  ~~~2~~& ~~~1~~&  ~~~ ~~& ~~~3~~& ~~~2~~& ~~~1~~& ~~~ ~~& ~~~3~~& ~~~2~~& ~~~4~~& ~~~6~~& ~~~3~~& ~~~2~~\\
~~~N$_{ev}$~~~	 &  ~~~1~~& ~~~3~~&  ~~~3~~& ~~~3~~& ~~~5~~& ~~~8~~& ~~~1~~& ~~~1~~& ~~~3~~& ~~~1~~& ~~~1~~& ~~~2~~& ~~~3~~\\

\hline\noalign{\smallskip}
\hline\noalign{\smallskip}
\end{tabular}
\end{table}

\begin{table}
\caption{\label{tab:4}  The distribution of 
\lq\lq white\rq\rq ~stars with respect to the charge topology in dissociation of $^{32}$S of 
the energy of 3.65 GeV per nucleon. }
\begin{tabular}{c|c|c|c|c|c|c|c|c|c|c|c|c|c|c|c|c|c|c}

\hline\noalign{\smallskip}
\hline\noalign{\smallskip}

~Z$_f$~	   &~ 15 ~&~ 14 ~&~ 14 ~&~ 13 ~&~ 13 ~&~ 12 ~&~ 12~&~ 11~&~ 11~&~ 10~&~ 10~&~ 10~&~~9 ~&~~8 ~&~~8 ~&~7+3&~~7 ~&5+3\\
~N$_{Z=1}$~&~~ 1 ~&~~   ~&~~ 2 ~&~~ 1 ~&~~ 3 ~&~~ 2 ~&~~ 4~&~~ 3~&~~ 5~&~~ 2~&~~ 4~&~~ 6~&~~3 ~&~~  ~&~~6 ~&~~4~&~~3 ~&~4\\
~N$_{Z=2}$~&~~   ~&~~ 1 ~&~~   ~&~~ 1 ~&~~   ~&~~ 1 ~&~~  ~&~~ 1~&~~  ~&~~ 2~&~~ 1~&~~  ~&~~2 ~&~~4 ~&~~1 ~&~~1~&~~3 ~&~2\\
~N$_{ev}$~ &~ 99 ~&~ 11 ~&~ 48 ~&~~ 7 ~&~~ 6 ~&~~ 3 ~&~~ 4~&~~ 4~&~~ 1~&~~ 1~&~~ 2~&~~ 1~&~~1 ~&~~1 ~&~~1 ~&~~1~&~~1 ~&~1\\

\hline\noalign{\smallskip}
\hline\noalign{\smallskip}
\end{tabular}
\end{table}

\indent  We also give the results obtained from an exposure involving $^{32}$S nuclei of the energy of 200 GeV per nucleon.
 In this case, the angular fragmentation cone is 0.5$^\circ$. In Table \ref{tab:4} the H isotope separation channel is seen to be 
dominant. In spite of poor statistics, a multifragmentation is revealed in the topology of 193  \lq\lq white\rq\rq ~stars. 
\par
\indent It is of interest to explore the  \lq\lq white\rq\rq ~star topology for heavy nuclei. Single events from a total
 breakup of Pb nuclei were observed in an emulsion exposed to ultra-relativistic Pb nuclei of the energy of 160 GeV (CERN).
 However it is impossible to perform a detailed study within the cone of the fragmentation of heavy nuclei even by the 
emulsion technique. It appears that this investigation can be carried out with the use of intense relativistic beams of 
heavy nuclei by measuring the total ionization and the energy fluxes in a total solid angle.   
\par

\section{\label{sec:level3} PARTICULAR FEATURES OF FRAGMENTATION OF 
B, C, N, AND O NUCLEI }
\subsection {Multifragmentation of $^{12}$C, $^{16}$O nuclei}
\indent The probabilities of formation of the systems consisting 
of a small number of fragments with Z=1 and 2 and their properties can be explored by means of a selection of 
\lq\lq white\rq\rq ~stars originating from the fragmentation of B, C, N, and O isotopes. Detailed information on the 
multifragmentation of the nuclei belonging to this group may be assumed as a basis for understanding processes occurring in
heavier nuclei. The dissociation of B and C nuclei to 3-body systems can proceed via the separation of the lightest nuclei,
 that is, $\alpha$ particles, deuterons, tritons, and  $^{3}$He nuclei, from the core in the form of an unstable $^{8}$Be
 nucleus, as well as via a direct fragmentation of them to  and He isotopes.
\par
\indent  The \lq\lq white\rq\rq ~stars from the $^{12}$C$^*\rightarrow$3$\alpha$ channel at the energy of 3.65 GeV per 
nucleon were studied in \cite{Belaga95,Belaga95a,Bondarenko98}. In particular, one demonstrated the role of the channel with a $^{8}$Be nucleus and 
one came to a conclusion about the transition to a direct multfragmentation with increasing total energy of a system 
consisting of three $\alpha$ particles. In \cite{Bogdanov86}, using the statistics 2757 inelastic interactions, it was established 
that no one event of binary splitting had been observed through the only possible $^{12}$C$^*\rightarrow^6$Li+$^6$Li
 channel.
\par

\begin{table}
\caption{\label{tab:5}  The distribution of 
\lq\lq white\rq\rq ~stars with respect to the charge topology in dissociation of $^{16}$0 of 
the energy of 3.65 GeV per nucleon. }

\begin{tabular}{c|c|c|c|c|c|c|c|c|c}
\hline\noalign{\smallskip}
\hline\noalign{\smallskip}

~~~Z$_f$~~~	     &  ~~~7~~&	~~~6~~&	~~~6~~&  ~~~5~~& ~~~5~~&  ~~~4~~& ~~~4~~& ~~~ ~~& ~~~ ~~\\
~~~N$_{Z=1}$~~~	 &  ~~~1~~& ~~~ ~~&	~~~2~~&  ~~~1~~& ~~~3~~&  ~~~2~~& ~~~ ~~& ~~~ ~~& ~~~2~~\\
~~~N$_{Z=2}$~~~	 &  ~~~ ~~&	~~~1~~&	~~~ ~~&  ~~~1~~& ~~~ ~~&  ~~~1~~& ~~~2~~& ~~~4~~& ~~~3~~\\
~~~N$_{ev}$~~~	 &  ~~18~~&	~~~21~~& ~~7~~&  ~~10~~& ~~~2~~&  ~~~1~~& ~~~1~~& ~~~9~~& ~~~3~~\\

\hline\noalign{\smallskip}
\hline\noalign{\smallskip}
\end{tabular}
\end{table}
\begin{table}
\caption{\label{tab:6}  The distribution of 
\lq\lq white\rq\rq ~stars with respect to the charge topology in dissociation of $^{16}$0 of 
the energy of 200 GeV per nucleon. }
\begin{tabular}{c|c|c|c|c|c|c|c|c|c|c|c}

\hline\noalign{\smallskip}
\hline\noalign{\smallskip}

~~~Z$_f$~~~	     &  ~~~7~~&	~~~6~~&	~~~6~~&  ~~~5~~& ~~~5~~&  ~~~4~~& ~~~3~~& ~~~3~~& ~~~ ~~& ~~~ ~~& ~~~ ~~\\
~~~N$_{Z=1}$~~~	 &  ~~~1~~& ~~~ ~~&	~~~2~~&  ~~~1~~& ~~~3~~&  ~~~2~~& ~~~1~~& ~~~3~~& ~~~ ~~& ~~~2~~& ~~~4~~\\
~~~N$_{Z=2}$~~~	 &  ~~~ ~~&	~~~1~~&	~~~ ~~&  ~~~1~~& ~~~2~~&  ~~~1~~& ~~~2~~& ~~~1~~& ~~~4~~& ~~~3~~& ~~~2~~\\
~~~N$_{ev}$~~~	 &  ~~49~~&	~~~6~~& ~~10~~&  ~~~5~~& ~~~1~~&  ~~~3~~& ~~~2~~& ~~~2~~& ~~~2~~& ~~~4~~& ~~~2~~\\

\hline\noalign{\smallskip}
\hline\noalign{\smallskip}
\end{tabular}
\end{table}
\indent In \cite{Avetyan96}, the \lq\lq white\rq\rq ~stars from the $^{16}$O$^*\rightarrow$4$\alpha$ channel were investigated using
 a large amount of information  (641 events). An analysis of the angular correlations gave evidence that the angular 
momentum was transferred to the systems of fragments and that the cascade decays via $^{8}$Be and $^{12}$C nuclei were
 nonessential. Tables \ref{tab:5} and \ref{tab:6} give the results of the selection of \lq\lq white\rq\rq ~stars using a sample of 2159 
interactions of $^{16}$O nuclei at the energy of 3.65 (72 stars) and at the energy of 200 GeV per nucleon (86 stars).  
\par
\subsection{Multifragmentation of $^{10}$B nucleus} \indent The study of the contribution from deuterons to the decays of odd-odd 
 $^{6}$Li \cite{Lepekhin95,Lepekhin98,Adamovich99,Adamovich03}, and $^{10}$B 
\cite{Bradnova03,Bradnova03a,Adamovich04}, and $^{14}$N nuclei pursues the investigation of the multifragmentation of light 
even-even nuclei with dissociation only to $\alpha$ particles. The role of the deuteron as a cluster is especially 
pronounced in the \lq\lq white\rq\rq ~stars of $^{6}$Li nuclei at the energy of 3.65 GeV per nucleon (in accordance with 
\cite{Adamovich99} 
 $^{6}$Li$^*\rightarrow$d+$\alpha$ - 74\%, $^{6}$Li$^*\rightarrow^3$Het$^*$ –- 13\%, $^{6}$Li$^*\rightarrow$tdp - 13\% ).    
\par\

\begin{table}
\caption{\label{tab:7}  The distribution of 
\lq\lq white\rq\rq ~stars with respect to the charge topology in dissociation of $^{10}$B of 
the energy of 1.0 GeV per nucleon. }

\begin{tabular}{c|c|c|c|c}
\hline\noalign{\smallskip}
\hline\noalign{\smallskip}

~~~Z$_f$~~~	     &  ~~~4~~~&	~~~3~~~&	~~~ ~~~&  ~~~ ~~~\\
~~~N$_{Z=1}$~~~	 &  ~~~1~~~& 	~~~ ~~~&	~~~3~~~&  ~~~1~~~\\
~~~N$_{Z=2}$~~~	 &  ~~~ ~~~&	~~~1~~~&	~~~1~~~&  ~~~2~~~\\
~~~N$_{ev}$~~~	 &  ~~~1~~~&	~~~5~~~&    ~~~5~~~&  ~~30~~~\\

\hline\noalign{\smallskip}
\hline\noalign{\smallskip}
\end{tabular}
%\hspace*{10cm}  % with the correct table height
\end{table}

\indent  The topology of \lq\lq white\rq\rq ~stars was investigated for $^{10}$B nuclei at the energy of 1.0 GeV per nucleon. 
Table \ref{tab:7} presents the charge topology distribution of 41 \lq\lq white\rq\rq ~stars with the angular cone for secondary
 tracks to 15$^\circ$ The fraction of the $^{10}$B$^*\rightarrow$d+$\alpha+\alpha$ decays is 40\% of the events with a
 charge topology 2+2+1. The contribution of the $^{10}$B$^*\rightarrow$d+$^{8}$Be+d$\rightarrow\alpha+\alpha$+d channel  
is estimated to be 18$\pm$3\%. The decay of an unstable $^{9}$B nucleus is not a basic source of the events with such a 
topology. This is suggested by the fact that the probability of observing a 4+1 topology in the
 $^{10}$B$^*\rightarrow$p+$^{9}$Be decay is small (3\%), as well as the contribution of $^{8}$Be to
 $^{10}$B$\rightarrow$p+$^{8}$Be is also not large. It may be concluded that the direct 3-body decays with
 \lq\lq white\rq\rq ~stars 2+2+1 configuration play a crucial role. Thus the decay topology  is indicative of an analogy 
with the $^{12}$C$^*\rightarrow$3$\alpha$ decay.
\par
\indent In order to gain extended knowledge about the relation between the direct 3-body decay and the decays via $^{8}$Be
 nucleus, emulsion was exposed to relativistic $^{9}$Be nuclei. A beam of $^{9}$Be nuclei with momentum of 2 GeV/c per
 nucleon was formed in $^{10}$B fragmentation after acceleration at the JINR Nuclotron. The process of production of 
\lq\lq white\rq\rq ~stars with 2 $\alpha$ particles is initiated in the fragmentation with a breakup of one neutron. 
An analysis of the data will allow one to have an idea about clustering in the $^{9}$Be nucleus and the probability of 
formation of  $^{8}$Be nucleus. This is expected to affect the yield of $\alpha$ particle pairs through n- $^{8}$Be and
 $\alpha$-n-$\alpha$ excitations. 
\par
\begin{table}
\caption{\label{tab:8}  The distribution of 
\lq\lq white\rq\rq ~stars with respect to the charge topology in dissociation of $^{14}$N of 
the energy of 2.1 GeV per nucleon. }

\begin{tabular}{c|c|c|c|c|c|c}
\hline\noalign{\smallskip}
\hline\noalign{\smallskip}

~~~Z$_f$~~~	     &  ~~~6~~~&	~~~5~~~&	~~~4~~~&  ~~~3~~~&	~~~ ~~~&  ~~~ ~~~\\
~~~N$_{Z=1}$~~~	 &  ~~~1~~~& 	~~~2~~~&	~~~1~~~&  ~~~4~~~&	~~~3~~~&  ~~~1~~~\\
~~~N$_{Z=2}$~~~	 &  ~~~ ~~~&	~~~ ~~~&	~~~1~~~&  ~~~ ~~~&	~~~2~~~&  ~~~3~~~\\
~~~N$_{ev}$~~~	 &  ~~~6~~~&	~~~3~~~&    ~~~1~~~&  ~~~1~~~&	~~~2~~~&  ~~12~~~\\

\hline\noalign{\smallskip}
\hline\noalign{\smallskip}
\end{tabular}
\end{table}
\subsection {Multifragmentation  of $^{14}$N nucleus} \indent It is interesting to find out the role of the 3-body decays which
 has been defined for $^{10}$B$^*\rightarrow$d+2$\alpha$, $^{12}$C$^*\rightarrow$3$\alpha$, and 
$^{16}$O$^*\rightarrow$4$\alpha$, as well as to develop ideas of clustering in nuclei involving deuterons. 
To this end, emulsion was exposed to $^{14}$N nuclei of the energy of 2.1 GeV per nucleon. The major goal is the study 
of the $^{14}$N$^*\rightarrow$d+3$\alpha$ \lq\lq white\rq\rq ~stars within the forward cone to 8$^\circ$. By 
the present time, data extracted from 540 interactions of $^{14}$N nuclei with the emulsion nuclei, including
25 \lq\lq white\rq\rq ~stars, were accumulated. Their distribution with respect to the charge topology is 
 given in Table \ref{tab:8}. There is an evidence for an important role of the 2+2+2+1  charge configuration, which is 
related to $^{14}$N decay. The noticeable role of the 6+1 configuration is seen to have analogy to the events 
with a Z=1 fragment splitting in the dissociation of heavier symmetric nuclei.
\par
\subsection {Clustering that involves tritons} \indent The study of the \lq\lq white\rq\rq ~stars of light odd-even 
stable nuclei ($^{7}$Li, $^{11}$B, $^{15}$N, and $^{19}$F) can provide a basis for including tritons into the 
general picture. It is established that in the \lq\lq white\rq\rq ~stars originating from relativistic 
$^{7}$Li nuclei, the $^{7}$Li$^*\rightarrow\alpha$t channel constitutes 50\%, 
$^{7}$Li$^*\rightarrow\alpha$+d+p - 30\%, and $^{7}$Li$^*\rightarrow\alpha$+p+2n - 20\% \cite{Adamovich04}. As a next step, 
an exposure has been performed and the dissociation of $^{11}$B nuclei of an the energy of 1.2 GeV is being 
analyzed. The major task of the experiment is the study of the \lq\lq white\rq\rq ~stars of
 the $^{11}$B$^*\rightarrow$t+2$\alpha$ channel.
\par

\section{\label{sec:level4} PROSPECTS OF THE STUDY OF 
NEUTRON-DEFICIENT Be, B, C, and N ISOTOPES }
\subsection {Search for a \lq\lq 3He process\rq\rq ~in $^{11}$C, $^{10}$C and $^{9}$C decays}
 \indent The $^{11}$B nucleus is a daughter one in 
the $\beta$ decay of a mirror $^{11}$C nucleus. Therefore following the study of the 
\lq\lq white\rq\rq ~stars of the  $^{11}$B$^*\rightarrow$t+2$\alpha$ and 
$^{11}$B$^*\rightarrow^7$Li+$\alpha$ channels it is interesting to explore the $^3$He role in
 $^{11}$C decays. The decays via the $^{11}$C$^*\rightarrow^3$He+2$\alpha$ and 
$^{11}$C$^*\rightarrow^7$Be+$\alpha$ channels may be analogous to those via the  
$^{12}$C$^*\rightarrow$3$\alpha$ and  $^{12}$C$^*\rightarrow^8$Be$\alpha$ channels. 
Clustering in $^{12}$C$^*\rightarrow$3$\alpha$ decays reflects the well-known 
\lq\lq 3$\alpha$ process\rq\rq ~in stars. Observation of the cluster
 $^{11}$C$^*\rightarrow^3$He+2$\alpha$ decays would serve as a basis for studying the possible role
 of the \lq\lq 3He process\rq\rq ~process  in nucleosynthesis in stars occurring by means
 of $^{11}$C$^*\leftarrow^3$He+2$\alpha$ fusion, that is, in helium media with a mixed 
composition of helium isotopes.   
\par
\indent  The $^{10}$C nucleus is produced out of a $^{9}$C nucleus by adding one neutron. 
However it appears that the addition of a neutron does not result in the formation in the ground 
state of $^{10}$C clusters in the form of deuterons or $^3$He nuclei. It is unlikely that the
2-cluster structures will be produced in the form of $^{7}$Be and $^{3}$He nuclei, or in the form
 of a $^{8}$B and a deuteron  because of a high binding energy of such clusters in the
 $^{10}$C nucleus. In the case of one external proton, an unstable $^{9}$B can serve as the
 central part of the nucleus. In the structure with two external protons, the central part 
is represented by another but also unstable $^{8}$Be nucleus. Such structures must apparently be
 similar to the Boromean  structures of neutron-excess nuclei. In the present case, one or two 
external protons keep the $^{10}$C nucleus from being decayed to nuclear resonant states.
\par
\indent  It is of interest to get experimental information about the channels 
 $^{10}$C$^*\rightarrow$+2$^3$He+$\alpha$ and $^{10}$C$^*\rightarrow^7$Be+$^3$He which permits one
 to make a generalization of the "3He process". In the above-mentioned irradiation of emulsion 
by the $^{10}$B nuclei, we have already observed two \lq\lq white\rq\rq ~stars from a dissociation
 without target-nucleus excitation that are interpreted as  
$^{10}$B$^*\rightarrow$t+$^3$He+$\alpha\rightarrow^{10}$Ñ$^*\rightarrow$+2$^3$He+$\alpha$.
They give an indication of the fact that there exists a $^{10}$C 3-cluster excitation mode. 
As an example, we note that the study of the t$\rightarrow^3$He charge exchange process on emulsion 
nuclei has shown a high reliability of its observation \cite{Adamovich03}. The $^{10}$C nucleus breakup can proceed 
in a cascade manner with the production in the intermediate state of $^{9}$B, $^{8}$Be and $^{6}$Be
 unstable nuclei with few charged fragments produced in the final state. Thus, it is possible to study
 the decays of these nuclei.  
\par
\indent  It is suggested to form at the JINR Nuclotron $^{11}$C and $^{10}$C  beams and expose to them emulsions.
 For the beam generation, one preferred the $^{11}$B$\rightarrow^{11}$C  and $^{10}$B$\rightarrow^{10}$C 
 charge exchange processes to the fragmentation  of heavier nuclei, so as to suppress the contribution from 
nuclei having close ionization.
\par
\indent Of all the nuclei being considered, the $^{9}$C nucleus has the largest ratio of the number of 
protons to that of neutrons. This nucleus has an additional proton with respect to the $^{8}$B nucleus. 
The binding energy of this proton is much higher than that of the external proton in $^{8}$B. Perhaps, this 
is an effect of the interaction of two protons, which is analogous to the interaction of external neutrons 
in $^{6}$He. Of special interest and urgency is the investigation of the probability of 
$^{9}$C$^*\rightarrow$3$^3$He decays with respect to $^{9}$C$^*\rightarrow$p+$^8$B, 2p+$^7$Be, and to some
 other decay channels. It should be noted that the larger the ratio Z/N in the nucleus being investigated 
thanks to more complete probability of observing nucleons from the fragmenting nucleus, the wider the
 manifestation of the advantages of the emulsion technique in the study of the \lq\lq white\rq\rq ~stars.
\par
\indent The fusion $^3$He+$^3$He+$^3$He$\rightarrow^6$Be+$^3$He$\rightarrow^9$C is one more option of
 the  \lq\lq 3He process\rq\rq. It's the $\beta$ decay to a mirror$^9$B nucleus that is not a bound one,
 results in the $^9$B$\rightarrow$p+2$\alpha$ decay. Thus, in a star medium, initially involving only the
 $^3$He, can proceed a workout of $^4$He. The $^9$C produced can take part in a further
 $^4$He$^9$Ñ$\rightarrow^{13}$N($\beta^+$)$\rightarrow^{13}$C fusion under definite astrophysical conditions.
\par
\subsection {Clusterung in $^8$B nucleus decays} \indent The particular feature of the $^8$B nucleus is a record low binding 
energy of one of the protons. Therefore, the $^8$B nucleus is most likely to have the core in the form of  a $^7$Be nucleus 
and a loosely bound proton the spatial distribution of which mostly determines the value of the $^8$B nucleus radius.    
\par
\indent The special features of the structure of light neutron-deficient nuclei may underlie the so-called 
{\it rp}-processes. For example, the presence of a state of the proton-halo type \cite{Schwab95} can positively affect 
the rate of synthesis of light radioactive nuclei along the boundary of proton stability that decay to stable isotopes.
 In particular, $^8$B halo reduces the Coulomb repulsion when p+$^3$He+$\alpha$ nuclei undergo a fusion in mixtures of
 the stable H and He isotopes in astrophysical systems. The $^8$B nucleus being produced can either  \lq\lq wait for\rq\rq 
~the $\beta$ decay or, in definite astrophysical scenarios, take part in fusion reactions 
$\alpha$+$^8$B$\rightarrow^{12}$N($\beta^+)\rightarrow^{12}$C. As compared with the $^{12}$C synthesis via the $^8$Be 
nucleus, this process features much longer life-time of the $^8$B nucleus.  
\par
\indent  The $^{10}$B nuclei with a momentum of 2.0 GeV/c per nucleon and an intensity of about 10$^8$ nuclei per cycle
 were accelerated at the JINR Nuclotron and a beam of secondary nuclei of a magnetic rigidity corresponding to Z/A = 5/8
 ($^{10}$B$\rightarrow^{8}$B fragmentation, as suggested in \cite{Schwab95}) was formed. Information on the $^{8}$B interactions 
in emulsion had been obtained. We plan to determine the probabilities of forming  \lq\lq white\rq\rq ~stars in 
$^8$B$\rightarrow^{7}$Be+p, p+$^3$He+$\alpha$, $^6$Li+2p, and $\alpha$+d+2p. In the $^8$B$\rightarrow^{7}$B
 fragmentation, a crossing of the limits of proton stability also takes place. Thus, there arises a possibility of 
studying the decay channels $^7$B$\rightarrow$p+2$^3$He (an analog to $^9$B) and 3p+$\alpha$. In order to investigate
 the  $^{12}$N structure and clear up the role played by  $^{8}$B in this nucleus it is intended to expose emulsion 
to the  $^{12}$N beam produced in the charge-exchange reaction  $^{12}$Ñ$\rightarrow^{12}$N. It is also possible to use
 the $^{12}$N$\rightarrow^{11}$N fragmentation to study decays of one more nucleus being away from the valley of proton
 stability.
\par
\subsection  {Clusterung in $^7$Be nucleus decays} \indent  The study of the $^7$Be nucleus fragmentation is of interest as far 
as this nucleus may be a core in the $^8$B nucleus. Using one and the same approach, it will be possible to compare the 
cluster structure of this nucleus with the $^6$Li \cite{Adamovich99} and $^7$Li \cite{Adamovich04} nuclei through the probabilities of forming
  \lq\lq white\rq\rq ~stars in the $\alpha$+$^3$He and p+$^6$Li channels.
\par

\begin{figure}
\includegraphics[width=160mm]{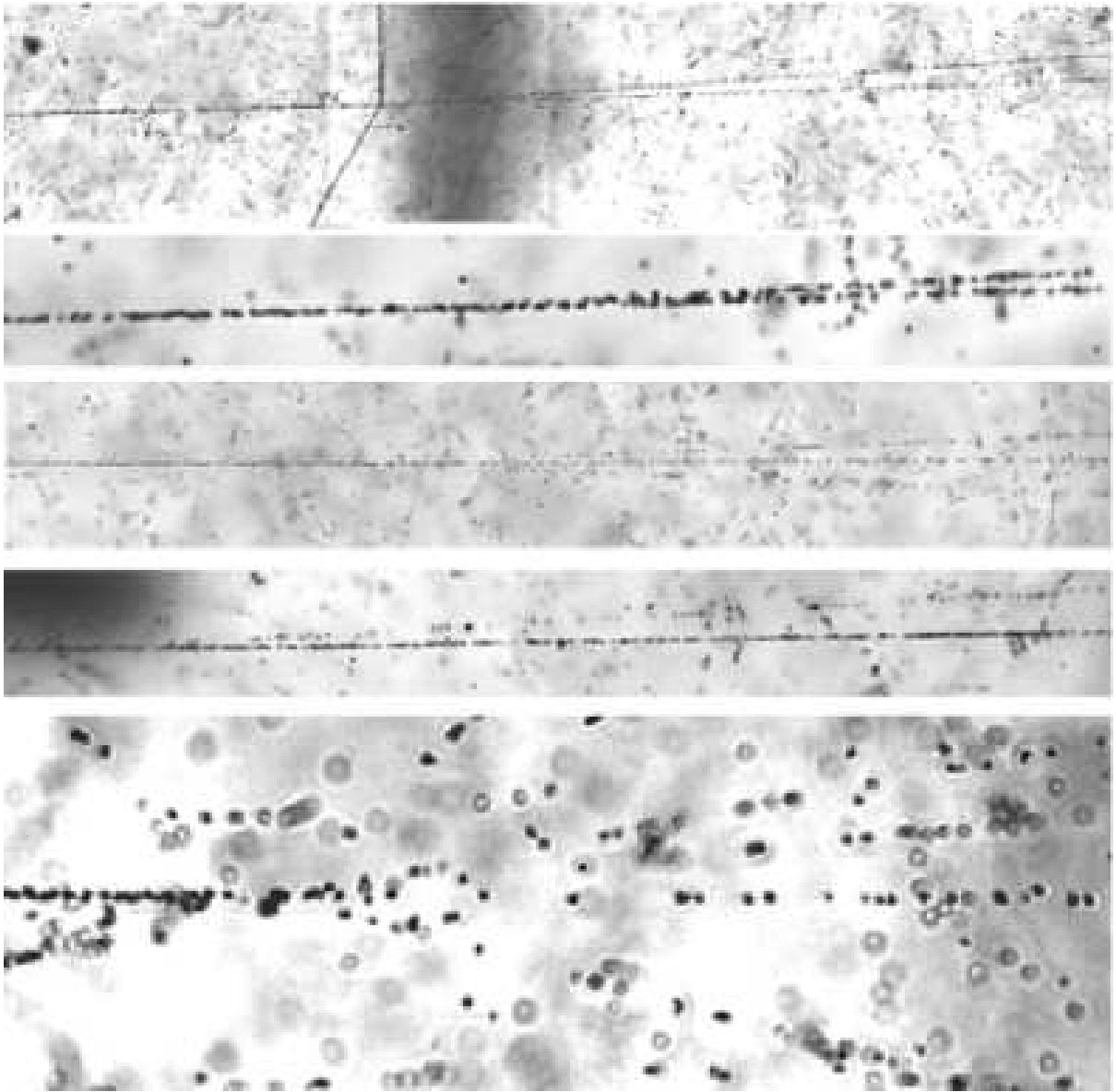} 
\caption{\label{fig:2} Examples of the events from the peripheral $^{7}$Be dissociation in emulsion. The upper photograph 
is a splitting to two He fragments with production of two target-nucleus fragments. Below there are 
\lq\lq white\rq\rq ~stars with splitting to two He, one He and two H, one Li and one H and four H fragments.}

\end{figure}
\indent  Emulsion was exposed to $^7$Be nuclei of the energy of 1.23 GeV per nucleon, the beam of which was formed at
 the JINR Nuclotron on the basis of the charge-exchange reaction $^7$Li$\rightarrow^{7}$Be. As a result of viewing over
 all the primary tracks, 75 \lq\lq white\rq\rq ~stars with the total secondary track charge equal to 4 were found in a 
cone up to 15$^\circ$.The examples of such stars for 2+2 topologies with and without target excitation, as well as for 
3+1 and 1+1+1+1 topologies are given in Figure \ref{fig:2}.
\par
\begin{table}
\caption{\label{tab:9}  The distribution of 
\lq\lq white\rq\rq ~stars with respect to the charge topology in dissociation of $^{7}$Be of 
the energy of 1.2 GeV per nucleon. }

\begin{tabular}{c|c|c|c|c}
\hline\noalign{\smallskip}
\hline\noalign{\smallskip}

~~~Z$_f$~~~	     &  ~~~3~~~&	~~~ ~~~&	~~~ ~~~&  ~~~ ~~~\\
~~~N$_{Z=1}$~~~	 &  ~~~1~~~& 	~~~4~~~&	~~~2~~~&  ~~~ ~~~\\
~~~N$_{Z=2}$~~~	 &  ~~~ ~~~&	~~~1~~~&	~~~1~~~&  ~~~2~~~\\
~~~N$_{ev}$~~~	 &  ~~~7~~~&	~~~2~~~&    ~~38~~~&  ~~28~~~\\

\hline\noalign{\smallskip}
\hline\noalign{\smallskip}
\end{tabular}
%\hspace*{10cm}  % with the correct table height
\end{table}
\indent  Table \ref{tab:9} shows the distribution of these stars over the charge topology channels. A channel with single-charged
 fragment splitting, which is unambiguously interpreted as p+$^6$Li, is observed. As a particular feature, it is possible
 to note two cases of a total breakup of the nucleus to singly charged fragments. In the case of 36 events with 2+1+1 
topology, 20 tracks with Z=2 were identified as $^3$He and 16 tracks - as $^4$He using the method of determination of 
the total momentum by means of multiple scattering. To separate the He nuclei according to their mass, use was made of 
a limiting value of the total momentum of P$\beta$c=5.1 GeV/c per nucleon fragments. To pursue further investigation it 
is of interest to analyze the $^7$Be$\rightarrow^{6}$Be(+n)$\rightarrow$2p+$^4$He+(n) channel, which is accompanied 
by the target-nucleus fragmentation initiated by a neutron.     
\par

\begin{figure}
\includegraphics[width=160mm]{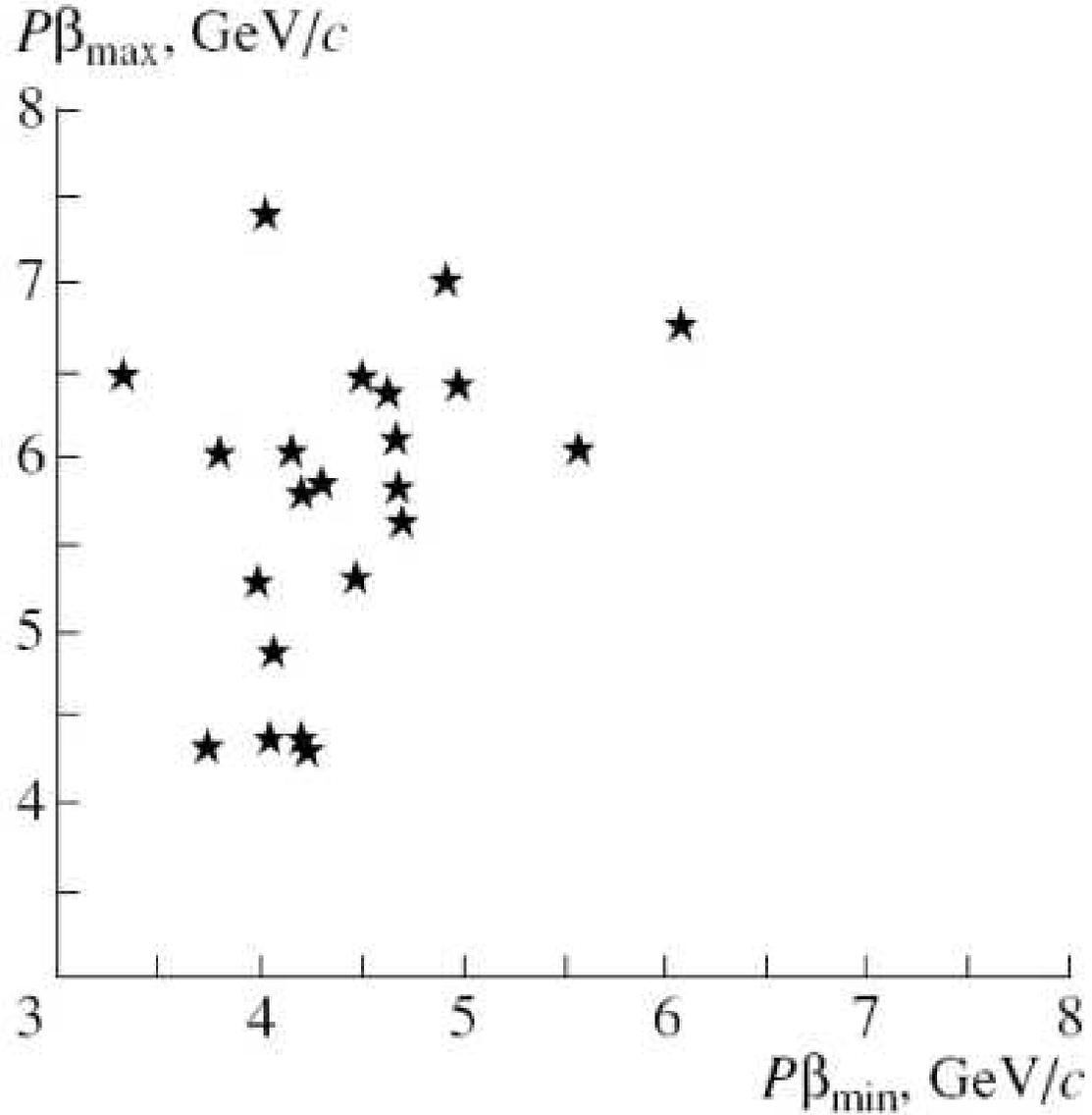} 
\caption{\label{fig:3} The distribution of the \lq\lq white\rq\rq ~stars from a $^{7}$Be nucleus of the energy of 1.23 GeV 
per nucleon with a decay to two He fragments with respect to minimum and maximum momenta. }

\end{figure}

\indent In Figure \ref{fig:3}, the 2-body decays are presented by points the coordinates of which are the total momenta P$\beta$c of
 fragments with Z= 2. The maximum P$\beta$c$_{max}$ value is attributed to the ordinate, and the minimum one P$\beta$c$_{min}$ to
 the abscissa. The distribution asymmetry is clearly seen. The $^7$Be$^*\rightarrow\alpha+^3$He decay, that occurs for a
 minimal excitation above   the decay threshold, as compared with other channels, is dominant in 22 events with 2+2
 topology. In the latter, 5 events are identified as the $^7$Be$^*(+n)\rightarrow^3$He+$^3$He decay. Thus, a clustering 
with $^3$He formation is clearly demonstrated in the \lq\lq white\rq\rq ~stars of the $^7$Be nucleus which makes it possible to put
the question as to whether this clustering is revealed in neighboring neutron-deficient nuclei. 
\par
\section{\label{sec:level5}Conclusions}
\indent  The experimental observations of the multifragmentation of light relativistic nuclei carried out by means of
 emulsions have been reviewed. Events of the \lq\lq white\rq\rq ~star type, which contain only tracks of the relativistic 
nucleus fragments, are selected. They involve neither charged meson tracks nor target-nucleus ones. The multifragmentation 
topology has been considered for these events.
\par
\indent  The characteristic feature of the charge topology in the fragmentation of Ne, Mg, Si and S nuclei implies an
 almost total suppression of pairing splitting of nuclei to fragments with charges larger than 2. Processes with separation
 of individual fragments occurring at minimal excitation energies are predominant. The growth of the nucleus fragmentation 
degree is revealed in an increase of the multiplicity of Z=1,2 fragments with decreasing charge of the non-excited part of
 the fragmenting nucleus.  
\par
\indent In multifragmentation processes of stable Li, Be, B, C, N and O isotopes special features of the formation of 
systems involving the lightest $\alpha$, d and t nuclei have been determined. In addition to the $\alpha$ clustering, 
a clustering of nucleons in the form of deuterons in $^{6}$Li and $^{10}$B decays, as well as of tritons in $^{7}$Li 
decays has been revealed. Besides, the multiparticle dissociation is found to be important for these nuclei. Emulsions 
exposed to relativistic $^{14}$N and $^{11}$B isotopes are being analyzed with the aim to study clustering of these types.
\par
\indent  The emergence of the $^{3}$He clustering can be detected in \lq\lq white\rq\rq ~stars, which is due to the
 neutron-deficient Be, B, C and N dissociation. Irradiation of emulsions by $^{7}$Be, $^{8}$B, and $^{9}$C nuclei has been 
performed. Irradiation by $^{10}$C, $^{11}$C, and $^{12}$N are planned. An analysis of the \lq\lq white\rq\rq ~stars from 
$^{7}$Be nuclei demonstrates the $^{3}$He clustering.
\par

\indent  Emulsions provide a unique basis for reconstructing relativistic multiparticle systems. Some of these systems are
 expected to play the role of the initial or intermediate loosely bound states in a fusion of more than two nuclei in 
nucleosynthesis in stars. The observation basis described in the paper can be employed in the search for such states.     
\par
\begin{acknowledgments}
\indent In conclusion, we would like to remember the names of our leaders in the domain of investigations with relativistic
 nuclei. Unfortunately, they are no more among the living. The foundations of the research  along these lines had been laid
 by Academician A.M.Baldin. For many years, M.I.Adamovich, V.I.Ostroumov, Z.I.Solovieva, K.D.Tolstov, M.I.Tretiakova, and 
G.M.Chernov had been leaders of the investigations carried out by nuclear emulsion technique at the JINR Synchrophasotron.
\par 
\indent  The results presented are based on a laborious visual search and measurements to which our laboratory assistants 
A.V. Pissetskaia (FIAN), L.N.Tkach (PINPh), N.A.Kachalova, I.I.Sosulnikova, A.M.Sosulnikova, and G.V. Stelmach (JINR) have 
made a valuable contribution.  I. I. Marjin (JINR) has ensured the maintenance of our microscopes. Emulsions have been
 processed with high quality by the chemical team  of the Laboratory of High Energies of JINR. A valuable contribution to
 our work has been given by the specialists of the Veksler and Baldin Laboratory of JINR ensured the nuclotron operation. 
We are grateful to the leaders of the Flerov Laboratory of Nuclear Reactions of JINR who has rendered support in urgent
 acquisition of emulsions.   
\par
\indent This work was supported by the Russian Foundation for Basic Research  ( project  nos.96-1596423, 
 02-02-164-12a, 03-02-16134, 03-02-17079 and  04-02-16593 ),  VEGA 1/9036/02  Grant from the Agency for Science 
of the Ministry for Education of the Slovak Republic and the Slovak Academy of Sciences, and grants from the JINR 
Plenipotentiaries of the Slovak Republic, the Czech Republic and Romania in 2002 and 2003.  
\par

\end{acknowledgments}

\end{document}